\documentclass{article}

\usepackage{amsfonts}
\usepackage{amsmath}
\usepackage{amssymb}

\newtheorem{defi}{Definition}[section]
\newtheorem{lemma}{Lemma}[section]
\newtheorem{theo}{Theorem}[section]

\newcounter{mnotecount}[section]
\renewcommand{\themnotecount}{\thesection.\arabic{mnotecount}}

\newcounter{mymnotecount}[section]
\renewcommand{\themymnotecount}{\thesection.\arabic{mymnotecount}}
\newcommand{\mymnote}[1]
{\protect{\stepcounter{mymnotecount}}$^{\mbox{\footnotesize $
\bullet$\themnotecount}}$ \marginpar{
\raggedright\tiny\em $\!\!\!\!\!\!\,\bullet$\themymnotecount: #1} }
\renewcommand{\mymnote}[1]{}

\begin{document}
\title{\normalsize \hfill UWThPh-2011-23\\
\normalsize \hfill July 2011\\[20mm]
\LARGE Existence and uniqueness of Bowen-York trumpets}
\author{G.~Waxenegger$^{1}$\thanks{Supported in part by Fonds zur F\"orderung der wissenschaftlichen Forschung in \"Osterreich, Projekt Nr. P20414-N16.
}, R.~Beig$^{1*}$ and N.~\'O Murchadha$^{2}$
\\*[3mm]
\small $^{1}$University of Vienna, Faculty of Physics, Gravitational Physics \\
\small Boltzmanngasse 5, 1090 Vienna, Austria \\
\small $^{2}$ Physics Department, University College Cork \\
\small Cork, Ireland}
\date{}
\maketitle

\begin{abstract}
We prove the existence of initial data sets which possess an asymptotically flat and an asymptotically cylindrical end. Such geometries are  known as trumpets in the community of numerical relativists.
\end{abstract}
\section{Introduction}
Initial data sets with one flat and one cylindrical asymptotic end - called trumpets in the recent numerical relativity literature -
are of great interest in current numerical black hole simulations based on the moving puncture method. In the present paper we give
an existence proof for classes of such data, which are vacuum, maximal, locally conformally flat and with extrinsic curvatures of the Bowen-York type. This problem has previously been solved
in \cite{dain} and \cite{gabach}, based on a method in which the cylindrical end is viewed as a certain singular limit of an asymptotically
flat end. We believe our work to be of independent interest, both for numerical and analytic purposes, since we completely avoid such a singular limit and since our method is basically a
straightforward version of the method of sub- and supersolutions. In particular we expect our results to extend to the case where the flat asymptotic end is replaced by a hyperboloidal asymptotic end.

Recall that an initial data set for the Einstein equations consists of a triple $\left(\bar{\Sigma},\bar{h}_{ij},\bar{K}_{ij}\right)$, where $\bar{\Sigma}$ is a $3$-dimensional manifold, $\bar{h}_{ij}$ a Riemannian metric and $\bar{K}_{ij}$ a symmetric tensor field on $\bar{\Sigma}$. Suppose there are given 'unphysical' quantities $(h_{ij},K_{ij})$, where $h_{ij}$ is a Riemannian metric on $\bar{\Sigma}$ and $K_{ij}$ is symmetric, trace and divergence free:
\begin{equation}
h^{ij}K_{ij}=0,\quad D^{i}K_{ij}=0.
\end{equation}
Then the quantities
\begin{equation}
\bar{h}_{ij}=\phi^{4}h_{ij},\quad \bar{K}_{ij}=\phi^{-2}K_{ij}
\end{equation}
satisfy the maximal initial-value constraints, namely
\begin{eqnarray}
\bar{D}^{j}\bar{K}_{ij}=0,\,\,\,\,\,\bar{K}=0,\\
\bar{R}-\bar{K}_{ij}\bar{K}^{ij}=0,
\end{eqnarray}
where $\bar{D}_{i}$ is the covariant derivative associated with $\bar{h}_{ij},\bar{R}$ the scalar curvature and $\bar{K}:=\bar{h}^{ij}\bar{K}_{ij}$,
provided that $\phi$ is everywhere positive and satisfies the Lichnerowicz equation given by
\begin{equation}\label{lich}
\triangle_{h}\phi-\frac{R}{8}\phi+\frac{1}{8}K_{ij}K^{ij}\phi^{-7}=0.
\end{equation}
We will be interested in the case where $\bar{\Sigma}$ is $\mathbb{R}^3 \setminus \{0\}$ equipped with the standard metric, i.e.,
$h_{ij} = \delta_{ij}$ in standard coordinates $x^i$. As for boundary conditions we require standard conditions of asymptotic flatness near infinity, which in our setting essentially amounts to $\phi = 1 + O(r^{-1})$  - which guarantees that $\bar{h}_{ij} - \delta_{ij} = O(r^{-1})$ - and
$K_{ij} = O(r^{-2})$ for large $r=|x|=(x^i x^j \delta_{ij})^{\frac{1}{2}}$. At the other (small $r$) end we want the physical metric $\bar{h}_{ij}$
to approach a cylindrical metric in the following sense: There should exist a diffeomorphism $\Phi$ between, say, $B_{1/2}\setminus\{0\}$
and $(T,\infty) \times S^2$ so that
\begin{equation}
(\Phi_{*}\bar{h})_{ij}-\tilde{h}_{ij}=o(1)\,\,\,\mathrm{as\,\,}t\,\,\rightarrow \infty\,,
\end{equation}
where $\tilde{h}$ denotes the cylindrical metric of the form
\begin{equation}\label{cyl}
\tilde{h}=dt^{2}+g_{\Xi}
\end{equation}
with $g_{\Xi}$ a Riemannian metric on the two sphere\footnote{We refrain from giving a definition - so far lacking in the literature - of an asymptotically cylindrical initial data set $(\bar{\Sigma}, \bar{h}_{ij}, \bar{K}_{ij})$, i.e., specifying general conditions on $\bar{K}_{ij}$.}.
Next note that, with $t = - \log r$, the manifold $(\mathbb{R}^3 \setminus \{0\}, r^{-2}\delta_{ij})$ is actually cylindrical with  $g_{\Xi}$ the standard metric on $S^2$. Thus the means, in our setting, to make the origin a cylindrical end
will be to demand that $\phi$ blows up like $r^{-\frac{1}{2}}$ for small $r$. An obvious way to guarantee this, as a simple scaling argument shows, is to require that $K_{ij}$ in (
\ref{lich}) blows up like
$r^{-3}$. A 7-parameter class of TT tensors satisfying the necessary conditions at both asymptotic ends are the following:\\
\begin{equation}\label{A}
K_{A}^{ij}=\frac{A}{r^{3}}\left(3n^{i}n^{j}-\delta^{ij}\right)\,,
\end{equation}
\begin{equation}\label{S}
K_{S}^{ij}=\frac{3}{r^{3}}\left(n^{i}\epsilon^{jkl}S_{k}n_{l}+n^{j}\epsilon^{ikl}S_{k}n_{l}\right)\,,
\end{equation}
where $(A,S_i)$ are constants, not all zero and
\begin{equation}\label{P}
K_{P}^{ij}=\frac{3}{2r^{2}}\left(P^{i}n^{j}+P^{j}n^{i}-(\delta^{ij}-n^{i}n^{j})P^{k}n_{k}\right).
\end{equation}
These are members of the Bowen-York \cite{by} class of TT tensors (see also \cite{beig}). The constants $S_i, P_i$ have the respective interpretation of total spin and total linear momentum of the configuration, measured at spatial infinity, which are conserved under time evolution. Furthermore $K_{A}^{ij}$ is the unique spherically symmetric $TT$ tensor on flat space.\\
The main result of our paper, here stated informally, is the following: Given $K^{ij} = K_{A}^{ij} + K_{S}^{ij} + K_P^{ij}$ with $(A,S_i)$ not all zero,
there is a unique solution of Eq.(\ref{lich}) with appropriate initial conditions so that $(\mathbb{R}^3 \setminus \{0\},\phi^4 \delta_{ij},
\phi^{-2}K_{ij})$ is an initial data set with one flat and one cylindrical asymptotic end. Interestingly, in the case of nonzero $S$,
the metric $g_\Xi$ on $S^2$, to which $g$ tends near the origin, is a non-constant multiple of the standard one, which is determined by a nonlinear equation on $S^2$.\\
When $S_i, P_i$ are zero, the solution to the above problem is known explicitly \cite{baumgarte} and by virtue of its spherical symmetry comes, of course, from a specific maximal slice of a Schwarzschild spacetime.  One is given $A$. One finds $M = \sqrt{\frac{4|A|}{3\sqrt{3}}}$ with $M > 0$ and $\phi_A$, which solves Eq.(\ref{lich}) given by
\begin{eqnarray}
\label{factor1}
\phi_A & = & \left[\frac{4R}{2R+M\left(4R^{2}+4M\,R+3M^{2}\right)^{1/2}}\right]^{1/2}\nonumber\\
& &\times\left[\frac{8R+6M+3\left(8R^{2}+8M\,R+6M^{2}\right)^{1/2}}{\left(4+3\sqrt{2}\right)\left(2R-3M\right)}\right]^{1/2 \sqrt{2}},
\end{eqnarray}
where the areal coordinate $R$ is implicitly defined as a function of $r$ as the inverse of the map $r(R)$ given by
\begin{eqnarray}\label{map}
r & = & \left[\frac{2R+M+(4R^{2}+4MR+3M^{2})^{1/2}}{4}\right]\nonumber\\
& &\times\left[\frac{(4+3\sqrt{2})(2R-3M)}{8R+6M+3(8R^{2}+8MR+6M^{2})^{1/2}}\right]^{1/\sqrt{2}}.
\end{eqnarray}
Note that the map (\ref{map}) is an orientation-preserving diffeomorphism from $(\frac{3 M}{2},\infty)$ to $(0,\infty)$.
Using this expression one easily obtains the asymptotic limits of the conformal factor
\begin{equation}
\label{limit1}
\phi_{A}\rightarrow \sqrt{\frac{3M}{2r}}\,(1 + O(r)) \textrm{\,\,as \,$r\rightarrow 0$}
\end{equation}
\begin{equation}
\label{limit2}
\phi_{A}\rightarrow 1+\frac{M}{2r} + O(r^{-2})\textrm{\,\,as $r\rightarrow\,\infty$}.
\end{equation}
Thus there is an asymptotically flat end for large $r$. Furthermore \eqref{limit1} ensures that small $r$ is an asymptotically cylindrical end where $r = e^{- \frac{2t}{3\!M}}$ and $g_{\Xi} = (\frac{3 M}{2})^2 d \Omega^2$, with $d \Omega^2$ the standard metric on $S^2$. Depending on the sign of $A$ in Eq.(8) the initial data set
$(\phi^4 \delta_{ij}, K^{ij}_A)$ corresponds to a slice in the
extended Schwarzschild spacetime which, when $A>0$, runs from spacelike
infinity to future timelike infinity, while if $A<0$ it goes to past timelike
infinity.\\
Now let $K^{ij}$ be given by $K^{ij}=K_{A}^{ij}+K_{S}^{ij} + K_P^{ij}$, defined respectively in (\ref{A}),(\ref{S}) and (\ref{P}),
with $(A,S_i)$ not all zero.  We try to solve  Eq.(\ref{lich}) with this form of extrinsic curvature. When we rotate the coordinates so that the angular momentum points in the $z$ direction we get
\begin{equation}
|K|^2=6\frac{A^{2}+3S^{2}\sin^{2}\theta}{r^{6}}+\frac{12A P^{i}n_{i}+18\epsilon_{ijz}n^{i}P^{j}S^{z}}{r^{5}}+\frac{9\left(P^{2}+2(P^{i}n_{i})^{2}\right)}{2r^{4}}
\end{equation}
 and find a trumpet solution. Since $K^{ij}_S$ has the same asymptotics as $K^{ij}_A$ we expect that our trumpet has the same asymptotics as the Baumgarte - Naculich \cite{baumgarte} trumpet. This means that the conformal factor should blow up at the origin like $1/\sqrt{r}$ (see (\ref{limit1})) and go like $1 + O(1/r)$ (see (\ref{limit2})) at infinity.\\
The outline of the paper is as follows: in Section (2) we first construct  weighted spaces with the desired asymptotic conditions. Then we show that a fairly general quasi-linear equation, which mimics the Lichnerowicz equation, has nice uniqueness and existence properties. The existence proof is an adaptation of the method of sub- and supersolutions. In Section (3) we are interested in a specific equation, the Lichnerowicz equation with Bowen-York sources, and show that it fits into the framework introduced in Section (2).
We begin with the pure angular momentum case. Here the Schwarzschild trumpet given by (\ref{factor1},\ref{map}) plays the role of the seed supersolution with a suitable choice of $A$. This gives us the desired existence and uniqueness result. Finally we show that the physical space goes to a true, but distorted, cylinder. We then extend the proof to include linear momentum.\\
We have taken a significant number of the details of the calculation and put them in an extensive appendix so as try and keep the flow of the main argument as simple as possible.

\section{The main theorem}
We use certain weighted H\"older spaces, which we now define:
\begin{defi}
Given $\alpha\in [0,1)$ and $\delta_{1} \in \mathbb{R}$, the space $C_{\delta_{1}}^{k,\alpha}\left(B_{3/2}\setminus\{0\}\right)$ is defined to be the set of functions $w\in C_{loc}^{k,\alpha}\left(B_{3/2}\setminus\{0\}\right)$ for which the following norm is finite
\begin{multline}
\label{norm1}
\left\|w\right\|_{C_{\delta_{1}}^{k,\alpha}\left(B_{3/2}\setminus\{0\}\right)}:=
\sum_{j=0}^k \sup_{|x| \in (0,\frac{3}{2})} |x|^{- \delta_1 + j}\,\left|\nabla^{j}w(x)\right|+\\
\sup_{|x| \in (0,\frac{3}{2})}\,\,\,\sup_{\substack{|y| \in [\frac{|x|}{2},|x|] \cap (0,\frac{3}{2}) \\x \neq y}}
|x|^{-\delta_1 + k + \alpha} \frac{\left|\nabla^{k}w\left(x\right)-\nabla^{k}w\left(y\right)\right|}{\left|x-y\right|^{\alpha}}.
\end{multline}
\end{defi}
Loosely speaking, a function in $C_{\delta}^{k,\alpha}\left(B_{3/2}\setminus\{0\}\right)$ is of $O(r^{\delta})$ for small $r$.
In particular for $\delta$ negative, which is the case of interest to us, this function may diverge at the origin no worse than $r^\delta$.\\
We also need spaces that dictate the behavior for $r\rightarrow\infty$:
\begin{defi}
For $k\in\mathbb{N}\cup\{0\}, \alpha\in [0,1)$ and $\delta_{1}\in\mathbb{R}$, the space $C_{\delta_{2}}^{k,\alpha}\left(\mathbb{R}^{3}\setminus B_{1}\right)$ is defined to be the set of functions $w\in C_{loc}^{k,\alpha}\left(\mathbb{R}^{3}\setminus B_{1}\right)$ for which the following norm is finite
\begin{multline}
\label{norm3}
\left\|w\right\|_{C_{\delta_{2}}^{k,\alpha}\left(\mathbb{R}^{3}\setminus B_{1}\right)} =  \sum_{j=0}^{k}\sup_{|x|\in [1,\infty)}|x|^{\delta_{2}+j}\left|\nabla^{j}w(x)\right| \\
+\sup_{|x|\in [1,\infty)}\,\,\sup_{|y|\in[\frac{|x|}{2},2 |x|\,]\cap \,[1,\infty)}|x|^{\delta_{2}+k+\alpha}\frac{\left|\nabla^{k}w\left(x\right)-\nabla^{k}w\left(y\right)\right|}{\left|x-y\right|^{\alpha}}.
\end{multline}
\end{defi}
Loosely speaking, a function in $C_{\delta}^{k,\alpha}\left(\mathbb{R}^{3}\setminus B_{1}\right)$ decays like $r^{- \delta}$ for large $r$,
when $\delta$ is positive.
Last but not least we use the norms \eqref{norm1}, \eqref{norm3} to define the weighted H\"older spaces that we will use in our results.
\begin{defi}
For $k\in\mathbb{N}\cup\{0\}, \alpha\in [0,1)$ and $\delta_{1},\delta_{2}\in\mathbb{R}$, the space $C_{\delta_{1},\delta_{2}}^{k,\alpha}\left(\mathbb{R}^{3}\setminus\{0\}\right)$ is defined to be the set of functions $w\in C_{loc}^{k,\alpha}\left(\mathbb{R}^{3}\setminus\{0\}\right)$ for which the following norm is finite
\begin{equation}
\left\|w\right\|_{C_{\delta_{1},\delta_{2}}^{k,\alpha}\left(\mathbb{R}^{3}\setminus\{0\}\right)}:=
\left\|w\right\|_{C_{\delta_{1}}^{k,\alpha}\left(B_{3/2}\setminus\{0\}\right)}+\left\|w\right\|_{C_{\delta_{2}}^{k,\alpha}\left(\mathbb{R}^{3}\setminus B_{1}\right)}.
\end{equation}
\end{defi}
In this section we prove the following:
\begin{theo}
Consider the equation
\begin{equation}
\label{equation1}
\triangle\phi=f(x,\phi)
\end{equation}
for a scalar function $\phi$ on $\mathbb{R}^{3}\setminus \left\{0\right\}$. Let $f\in C^{1}(\mathbb{R}^{3}\setminus\left\{0\right\}\times\mathbb{R}^{+})$ be non-decreasing in the second argument and non-positive. Assume that there exist positive functions $\phi_{+}$ and $\phi_-$, such that (here and subsequently we assume that $0 < \epsilon < \frac{1}{2}$)
\begin{itemize}
\item[(i)] $\phi_{\pm}-1\in C_{-1/2,1/2+\epsilon}^{2,\alpha}\left(\mathbb{R}^{3}\setminus \left\{0\right\}\right),\,0 < \alpha < 1$,
\item[(ii)] $\phi_- \leq \phi_+$
\end{itemize}
and
\begin{itemize}
\item[(iii)] $\triangle\phi_{+}\leq f(x,\phi_{+})\,\,,\,\,\triangle\phi_-\geq f(x,\phi_-)$.
\end{itemize}
For all functions $\phi$ with $\phi - 1 \in C_{-1/2,1/2+\epsilon}^{0,\alpha}\left(\mathbb{R}^{3}\setminus\left\{0\right\}\right),\,0 \leq \alpha < 1$, that are pointwise bounded between $\phi_{-}$ and $\phi_{+}$, i.e., $\phi_{-}\leq\phi\leq\phi_{+}$, we have that
\begin{itemize}
\item[(iv)]
$f(x,\phi)\in C_{-5/2,5/2+\epsilon}^{0,\alpha}\left(\mathbb{R}^{3}\setminus\left\{0\right\}\right)$
\end{itemize}
and
\begin{itemize}
\item[(v)] $\frac{\partial f}{\partial\phi}(x,\phi)\in C_{-2,2}^{0,\alpha}\left(\mathbb{R}^{3}\setminus\left\{0\right\}\right)$.
\end{itemize}
Then there exists a unique solution $\phi '$ of \eqref{equation1} such that
\begin{itemize}
\item[(a)] $\phi '-1\in C_{-1/2,1/2+\epsilon}^{2,\alpha}\left(\mathbb{R}^{3}\setminus\left\{0\right\}\right),\,0 < \alpha < 1$,
\end{itemize}
and
\begin{itemize}
\item[(b)] $\phi_{-}\leq\phi '\leq\phi_{+}$.
\end{itemize}
\end{theo}
\emph{Proof.} Before starting, we point out the following fact, which is easily verified: Let $w$ be any function which belongs to $C_{-2,2}^{0,\alpha}\left(\mathbb{R}^{3}\setminus \left\{0\right\}\right)$ and let the function $g\in C_{-2,2}^{0,\alpha}\left(\mathbb{R}^{3}\setminus \left\{0\right\}\right)$ be given by
\begin{equation}
g\left(x\right):=\frac{c}{r^{2}},
\end{equation}
where $c$ is a positive number. Then one can always choose the constant $c$ large enough so that $g\geq |w|$
pointwise. We choose $c$ large enough so that
\begin{equation}
\frac{c}{r^2}\geq|\frac{\partial f}{\partial\phi}|\,\,\,\,\mathrm{in}\,\mathbb{R}^3 \setminus \{0\}\,\,\,\,\, \mathrm{for\,all}\,\,\phi_- \leq \phi \leq \phi_+.
\end{equation}
We will now construct a pointwise increasing sequence of functions, starting from $\phi_{-}$. We solve
\begin{equation}\label{one}
(\triangle - g)\phi_1 = f(x,\phi_-) - g \phi_- \,\,\,\Longleftrightarrow\,\,\,
(\triangle - g) u_{1}=f\left(x,\phi_{-}\right)-g\left(\phi_{-}-1\right)\,
\end{equation}
with $\phi_1 = 1 + u_1$. Eq.(\ref{one}) has a unique solution $u_{1}\in C_{-1/2,1/2+\epsilon}^{2,\alpha}\left(\mathbb{R}^{3}\setminus \left\{0\right\}\right)$, since the right hand side of the equation for $u_1$ is in $C_{-5/2,5/2+\epsilon}^{0,\alpha}\left(\mathbb{R}^{3}\setminus \left\{0\right\}\right)$ and the operator is an isomorphism between these function spaces (see Theorem A.1). From the equations satisfied respectively by $\phi_{1}$ and $\phi_{-}$ we deduce the following inequality:
\begin{equation}
(\triangle - g)\left(\phi_{1}-\phi_{-}\right)=f\left(x,\phi_{-}\right)- \triangle\phi_{-} -g\phi_- + g\phi_- \leq 0.
\end{equation}
Hence, from Lemma A.1,
\begin{equation}
\phi_{1}\geq\phi_{-}.
\end{equation}
Also,
\begin{multline}
(\triangle - g)(\phi_{+}-\phi_{1})= \triangle \phi_+ - f(x,\phi_-) - g(\phi_+ - \phi_-) \leq \\
\leq f\left(x,\phi_{+}\right)-f\left(x,\phi_{-}\right)-g\left(\phi_{+}-\phi_{-}\right)=\\
=\left(\phi_{+}-\phi_{-}\right)\int_{0}^{1}\{\frac{\partial f}{\partial\phi}\left(x,\phi_{-}+t\left(\phi_{+}-\phi_{-}\right)\right)-g\}dt.
\end{multline}
In particular
\begin{equation}
\triangle\left(\phi_{+}-\phi_{1}\right)-g\left(\phi_{+}-\phi_{1}\right)\leq 0\,,
\end{equation}
whence $\phi_{1}\leq\phi_{+}$. We define
\begin{equation}
u_{n}\equiv\phi_{n}-1
\end{equation}
and use the following induction formula:
\begin{equation}
(\triangle - g)\phi_n = f(x,\phi_{n-1}) - g \phi_{n-1} \,\,\,\Longleftrightarrow\,\,\, (\triangle - g) u_{n}=f\left(x,\phi_{n-1}\right)-gu_{n-1}.
\end{equation}
We suppose that $\phi_{m}$ exists for $0\leq m\leq n-1$ with $\phi_{0}=\phi_{-}$ and $u_{m}\in C_{-1/2,1/2+\epsilon}^{2,\alpha}\left(\mathbb{R}^{3}\setminus \left\{0\right\}\right)$ for $0\leq m\leq n-1$ and that for these
\begin{equation}
\phi_{-}\leq\phi_{m-1}\leq\phi_{m}\leq\phi_{+}.
\end{equation}
Theorem (A.1) from the Appendix shows that $u_{n}\in C_{-1/2,1/2+\epsilon}^{2,\alpha}\left(\mathbb{R}^{3}\setminus \left\{0\right\}\right)$ exists.
By arguments completely analogous to that for $\phi_1$ we find that
\begin{equation}
\phi_{n-1} \leq \phi_n \leq \phi_+.
\end{equation}
Thus $\phi_{n}$ converges at each point to a limit $\phi\left(x\right)=1+u\left(x\right)$, with
$\phi_{-}\leq\phi\leq\phi_{+}$. It turns out that $\phi$ is a solution of our original equation. The sequence $\{u_{n}\}$ is uniformly bounded in the $C_{-1/2,1/2+\epsilon}^{2,\alpha}\left(\mathbb{R}^{3}\setminus \left\{0\right\}\right)$ norm and $C_{-1/2,1/2+\epsilon}^{2,\alpha}\left(\mathbb{R}^{3}\setminus \left\{0\right\}\right)$ is compactly embedded in $C_{-1/2-\epsilon,1/2}^{2,\alpha '}\left(\mathbb{R}^{3}\setminus \left\{0\right\}\right)$, where $\alpha '<\alpha$ (see Lemma (A.3) of the Appendix). Hence there is a subsequence, still denoted $u_{n}$, which converges in the $C_{-1/2-\epsilon,1/2}^{2,\alpha '}\left(\mathbb{R}^{3}\setminus \left\{0\right\}\right)$ norm to a function $u\in C_{-1/2-\epsilon,1/2}^{2,\alpha '}\left(\mathbb{R}^{3}\setminus \left\{0\right\}\right)$, identical to the previously defined $u$. Later we will show that $u\in C_{-1/2,1/2+\epsilon}^{2,\alpha}\left(\mathbb{R}^{3}\setminus \left\{0\right\}\right)$. The functions $f\left(\cdot,\phi_{n}\right)$ converge to $f\left(\cdot,\phi\right)$ in the $C_{-5/2-\epsilon,5/2}^{0,\alpha}\left(\mathbb{R}^{3}\setminus \left\{0\right\}\right)$ norm because of the following inequality
\begin{equation}\label{cont}
\|f\left(\cdot,\phi\right)-f\left(\cdot,\phi_{n}\right)\|_{C_{-5/2-\epsilon,5/2}^{0,\alpha}}\leq C\|\phi-\phi_{n}\|_{C_{-1/2-\epsilon,1/2}^{0,\alpha}}\,,
\end{equation}
which results form combining the mean value theorem with the multiplication lemma of the Appendix.
By continuity of the respective maps we have that $(\triangle - g)u_n$ converges to $(\triangle - g)u$ and $F(\cdot,u_n)$ converges to $F(\cdot,u)$,
where $F(\cdot,u)=f(\cdot,u+1) - g u$, both in the $C_{-5/2-\epsilon,5/2}^{0,\alpha}$ norm (see (\ref{cont}) for the latter fact). But
\begin{multline}
\lim_{n\rightarrow\infty}\|(\triangle - g)u_{n}-F\left(\cdot,u_{n}\right)\|_{C_{-5/2-\epsilon,5/2}^{0,\alpha}}=\\
= \lim_{n\rightarrow\infty}\|F\left(\cdot,u_{n-1}\right)-F\left(\cdot,u_{n}\right)\|_{C_{-5/2-\epsilon,5/2}^{0,\alpha}} = 0.
\end{multline}
Since $\phi_- \leq \phi \leq \phi_+$, we also have that
$u=\phi-1\in C_{-1/2,1/2+\epsilon}^{0}\left(\mathbb{R}^{3}\setminus \left\{0\right\}\right)$.
Furthermore the equation $(\triangle - g)u=F\left(x,u\right)$ shows that
\begin{equation}
\triangle u=f(x,\phi).
\end{equation}
Hence
\begin{equation}
\label{equation3}
u(x)= - \frac{1}{4\pi}\int\frac{f(x',\phi)}{|x-x'|}dx',
\end{equation}
where
\begin{equation}
f(\cdot,\phi)\in C_{-5/2,5/2+\epsilon}^{0}\left(\mathbb{R}^{3}\setminus \left\{0\right\}\right).
\end{equation}
By directly estimating the Poisson integral \eqref{equation3} we get
\begin{equation}
\nabla u\in C_{-3/2,3/2+\epsilon}^{0}\left(\mathbb{R}^{3}\setminus \left\{0\right\}\right)\,
\end{equation}
which implies
\begin{equation}
\label{equation4}
u\in C_{-1/2,1/2+\epsilon}^{0,\alpha}\left(\mathbb{R}^{3}\setminus \left\{0\right\}\right)\,,
\end{equation}
where we are using the elementary fact that $w$ is an element of some $C_{\delta_{1}}^{0,\alpha}(\mathbb{R}^3 \setminus B_{3/2})$, when locally $r(x)^{-\delta_{1}}|w(x)|\leq C$ and $r(x)^{-\delta_{1}+1}|\nabla w(x)|\leq C$ and a function $w$ is an element of
$C_{\delta_{2}}^{0,\alpha}(\mathbb{R}^3 \setminus B_1)$, when locally $r(x)^{\delta_{2}}|w(x)|\leq C$ and $r(x)^{\delta_{2}+1}|\nabla w(x)|\leq C$.
Using \eqref{equation4}, the definition of $f$, and Lemma (A.3) gives
\begin{equation}
F\left(\cdot,u\right)\in C_{-5/2,5/2+\epsilon}^{0,\alpha}\left(\mathbb{R}^{3}\setminus \left\{0\right\}\right).
\end{equation}
From the isomorphism property of $\triangle - g$ and the bounded inverse theorem of linear functional analysis we infer that
\begin{equation}
\|w\|_{C_{-1/2,1/2+\epsilon}^{2,\alpha}}\leq C\|(\triangle - g)w\|_{C_{-5/2,5/2+\epsilon}^{0,\alpha}}.
\end{equation}
Thus
\begin{equation}
u=\phi-1\in C_{-1/2,1/2+\epsilon}^{2,\alpha}\left(\mathbb{R}^{3}\setminus \left\{0\right\}\right).
\end{equation}
As for uniqueness, suppose there are two solutions, $\phi_{1}$ and $\phi_{2}$, which fulfill
\begin{equation}
\triangle\phi_{1}=f(x,\phi_{1})\,,\hspace{0.4cm}\triangle\phi_{2}=f(x,\phi_{2})
\end{equation}
subject to our boundary conditions. Let us consider the following function
\begin{equation}
\sigma:=r^{1/2+\epsilon/2}\left(\phi_{1}-\phi_{2}\right).
\end{equation}
The function $\sigma$ tends to $0$ at the origin and at infinity. We now compute $\triangle \sigma$ (see (A.1)).
Assuming $\sigma$ to have a positive maximum or a negative minimum and using the non-decreasing nature of $f(\cdot,\phi)$, this yields a contradiction, which ends the proof of Theorem (2.1).
\section{Bowen York sources}
\begin{theo}\label{function}
Let $K^{ij}$ be given by $K^{ij}=K_{A}^{ij}+K_{S}^{ij}$, defined respectively in (\ref{A}) and (\ref{S}),
with $(A,S_i)$ not all zero.
Then, there exists a unique positive $\phi$ with the property that $\phi-1\in C_{-1/2,1/2+\epsilon}^{2,\alpha}\left(\mathbb{R}^{3}\setminus\left\{0\right\}\right)$ and which is a solution of
\begin{equation}
\triangle\phi=-\frac{1}{8}K_{ij}K^{ij}\phi^{-7}=:f(\cdot,\phi).
\end{equation}
\end{theo}
\emph{Proof.} First note that
\begin{equation}
\label{equation6}
K_{ij} K^{ij} = |K|^2 = |K_A|^2 + |K_S|^2 = 6\frac{A^{2}+3S^{2}\sin^{2}\theta}{r^{6}},
\end{equation}
where $\theta$ is the angle between the angular momentum vector $\vec{S} = (S^x, S^y, S^z)$ and the radial vector $\vec{R} = (x, y, z)$. Alternatively, if we rotate the coordinates so that the angular momentum is along the $z$ axis, $\theta$ is the standard inclination angle. Clearly $f$ is in $C^{1}(\mathbb{R}^{3}\setminus\left\{0\right\}\times\mathbb{R}^{+})$,  is non-decreasing in the second argument and non-positive. Recall that $\phi_A$ referred to in the introduction in Eqs.(\ref{factor1}, \ref{map}) satisfies
$\triangle \phi_A = - \frac{1}{8} |K_A|^2 \phi_A^{-7}$. Now choose $A'$ large enough so that $A'^2 > A^2 + 3S^2$. This guarantees $|K_{A'}|^2 \geq |K|^2$. It follows that $\phi_{A'}$ furnishes a supersolution, again called $\phi_{+}$. Next we observe that $\phi_{+}-1$ belongs to $C_{-1/2,1/2+\epsilon}^{2,\alpha}\left(\mathbb{R}^{3}\setminus\left\{0\right\}\right)$. Furthermore
\begin{equation}
 K_{ij}K^{ij}\phi_{+}^{-7}=6\frac{A^{2}+3S^{2}\sin^{2}\theta}{r^{6}}\phi_{+}^{-7}\in C_{-5/2,5/2+\epsilon}^{0,\alpha}\left(\mathbb{R}^{3}\setminus \left\{0\right\}\right).
\end{equation}
Here we are making crucial use of the asymptotic behaviour of $\phi_+$ near the origin, in particular that
$r^\frac{1}{2} \phi_+$ tends to a positive constant as $r$ goes to zero.
Next we define $\phi_{-}$ by setting $\phi_{-}:=u_{-}+1$, where $u_{-}$ fulfills
\begin{equation}\label{sub}
\triangle u_{-}=-\frac{1}{8}K_{ij}K^{ij}\phi_{+}^{-7}.
\end{equation}
From the Poisson integral or Theorem A.1, there exists $u_-$ satisfying (\ref{sub}) in $C_{-1/2,1/2+\epsilon}^{2,\alpha}\left(\mathbb{R}^{3}\setminus\left\{0\right\}\right)$ which, by Lemma A.1 is non-negative.
Now calculate
\begin{equation}
\triangle(\phi_+ -\phi_-)=\triangle\phi_+ -f(x,\phi_+)\leq 0.
\end{equation}
Hence, from Lemma (A.1), we have $\phi_-\leq\phi_+$. From the non-decreasing property of the function $f$ it follows that $\phi_{-}$ furnishes a subsolution. In addition the Poisson integral shows that, like $\phi_+$, the function $r^\frac{1}{2}\phi_-$
has a positive limit at the origin and
\begin{equation}
 K_{ij}K^{ij}\phi_-^{-7}=6\frac{A^{2}+3S^{2}\sin^{2}\theta}{r^{6}}\phi_-^{-7}\in C_{-5/2,5/2+\epsilon}^{0,\alpha}\left(\mathbb{R}^{3}\setminus \left\{0\right\}\right)
\end{equation}
and
\begin{equation}
\frac{\partial f}{\partial \phi}(x,\phi_-)\propto K_{ij}K^{ij}\phi_-^{-8}=6\frac{A^{2}+3S^{2}\sin^{2}\theta}{r^{6}}\phi_-^{-8}\in C_{-2,2}^{0,\alpha}\left(\mathbb{R}^{3}\setminus \left\{0\right\}\right).
\end{equation}
Similarly we check the validity of assumptions $(iv)$ and $(v)$ for all $\phi$'s with $\phi_- \leq \phi \leq \phi_+$.
This ends the proof of the Theorem.\\
More generally we have the following result:
\begin{theo}
\label{functiona}
Let $K^{ij}$ be given by $K^{ij}=K_{A}^{ij}+K_{S}^{ij}+ K_P^{ij}$, defined respectively in (\ref{A},\ref{S},\ref{P}),
with $(A,S_i)$ not all zero.
Then, there exists a unique positive $\phi$ with the property that $\phi-1\in C_{-1/2,1/2+\epsilon}^{2,\alpha}\left(\mathbb{R}^{3}\setminus\left\{0\right\}\right)$ and which is a solution of
\begin{equation}
\label{equation21}
\triangle\phi=-\frac{1}{8}K_{ij}K^{ij}\phi^{-7}=:f(\cdot,\phi).
\end{equation}
\end{theo}
\emph{Proof.} Here it is simplest to use, similar as in \cite{gabach}, as supersolution the conformal factor
coming from the extreme Reissner Nordstr\"om trumpet initial data, namely
\begin{equation}
\phi_{+}=\sqrt{1+\frac{Q}{r}}.
\end{equation}
Then, choosing
\begin{equation}
Q^{2}\geq 3A+9S+7P^{2},
\end{equation}
we have that
\begin{equation}
\triangle\phi_{+}=-\frac{1}{4}\frac{Q^{2}}{r^{4}}\phi_{+}^{-3}\leq-\frac{1}{8}|K|^{2}\phi_{+}^{-7}.
\end{equation}
Clearly there again holds that
\begin{equation}
\phi_{+}-1\in C_{-1/2,1/2+\epsilon}^{2,\alpha}\left(\mathbb{R}^{3}\setminus\left\{0\right\}\right).
\end{equation}
The rest of the argument follows as in the proof of the previous theorem.\\

The importance of the previous results lies in the fact that now we are able to show the following lemma.
\begin{lemma}
Let $\phi$ be a solution of \eqref{equation21}, such that $\phi-1\in C_{-1/2,1/2+\epsilon}^{2,\alpha}\left(\mathbb{R}^{3}\setminus\left\{0\right\}\right)$. Define
\begin{equation}
\bar{h}_{ij}=\phi^{4}\delta_{ij},\quad \bar{K}_{ij}=\phi^{-2}K_{ij}.
\end{equation}
The resulting initial data set $\left(\mathbb{R}^{3}\setminus\left\{0\right\},\bar{h}_{ij},\bar{K}_{ij}\right)$ has an asymptotically flat and an asymptotically cylindrical end.
\end{lemma}
\emph{Proof.} First of all we show that $u=\phi-1\in C_{-1/2,1/2+\epsilon}^{2,\alpha}\left(\mathbb{R}^{3}\setminus\left\{0\right\}\right)$ can be uniquely decomposed as
\begin{equation}
u=\frac{d(\theta,\varphi)}{\sqrt{r}}\chi(r)+h,
\end{equation}
where $h$ diverges more slowly at the origin than $r^{-\frac{1}{2}}$ and  $\chi(r)$ is a bump function with support in $|x|\leq 2$ and which equals $1$ on $|x|\leq 1$. The quantity $d$ is a $C^{\infty}$ function on the $2$-sphere, which is subject to the following nonlinear equation
\begin{equation}
\label{equation5}
\left(\triangle_{S^{2}}-\frac{1}{4}\right)d=-\frac{1}{8}\{6A^{2}+18S^{2}\sin^{2}\theta\}d^{-7}.
\end{equation}
The origin of the nonlinear equation for $d$ can be found \cite{hannam2}, but will of course be implicit in the ensuing proof. For an existence proof for a unique $d$, which is also positive, the reader is referred to \cite{gabach}.
Given $d$, define $h$ as the solution to equation
\begin{equation}
\label{equation7}
\triangle h=\triangle\left(u-\frac{d}{\sqrt{r}}\chi(r)\right).
\end{equation}
The function  $h$ is clearly in $C_{-1/2,1/2+\epsilon}^{2,\alpha}\left(\mathbb{R}^{3}\setminus\left\{0\right\}\right)$.

But we want to show more. Using \eqref{equation6} and \eqref{equation7},  we calculate
\begin{eqnarray}
\label{equation8}
\triangle h & = & - \frac{1}{8}K_{ij}K^{ij}(u+1)^{-7}-\triangle\left(\frac{d}{\sqrt{r}}\right)\chi(r)\nonumber\\
& & -2\vec{\nabla}\left(\frac{d}{\sqrt{r}}\right)\cdot\vec{\nabla}\chi(r)-\frac{d}{\sqrt{r}}\triangle\chi(r)\nonumber\\
& = & \frac{1}{8}K_{ij}K^{ij}\left(\left(\frac{d}{\sqrt{r}}\right)^{-7}\chi(r)-\left(\frac{d}{\sqrt{r}}\chi(r)+h+1\right)^{-7}\right)\nonumber\\
& & -2\vec{\nabla}\left(\frac{d}{\sqrt{r}}\right)\cdot\vec{\nabla}\chi(r)-\frac{d}{\sqrt{r}}\triangle\chi(r),
\end{eqnarray}
where in the second equality sign we have used that the curly bracket on the r.h. side of (\ref{equation5}) is equal to
$K_{ij} K^{ij} r^6$.
With the help of the function
\begin{equation}
\rho(r):=\chi(r)^{-1/7},
\end{equation}
we can write \eqref{equation8} as
\begin{eqnarray}
\triangle h & = & \frac{1}{8}K_{ij}K^{ij}\left(h+1+\left(\chi(r)-\rho(r)\right)\frac{d}{\sqrt{r}}\right)\sum_{i=0}^{6}\left(\frac{d}{\sqrt{r}}\rho(r)\right)^{i-7}\phi^{-1-i}\nonumber\\
& & -2\vec{\nabla}\left(\frac{d}{\sqrt{r}}\right)\cdot\vec{\nabla}\chi(r)-\frac{d}{\sqrt{r}}\triangle\chi(r),
\end{eqnarray}
where we have used the following elementary identity
\begin{equation}
\frac{1}{a^{p}}-\frac{1}{b^{p}}=\left(b-a\right)\sum_{i=0}^{p-1}a^{i-p}b^{-1-i},
\end{equation}
which is true for real numbers $a$ and $b$. We define
\begin{equation}
c(r,\theta,\varphi):=\frac{r^{2}}{8}K_{ij}K^{ij}\left(\sum_{i=0}^{6}\left(\frac{d}{\sqrt{r}}\rho(r)\right)^{i-7}\phi^{-1-i}\right)
\end{equation}
and furthermore
\begin{equation}
f(r,\theta,\varphi):=\frac{c(r,\theta,\varphi)}{r^{2}}\left(1+\left(\chi(r)-\rho(r)\right)\frac{d}{\sqrt{r}}\right)-2\vec{\nabla}\left(\frac{d}{\sqrt{r}}\right)\cdot\vec{\nabla}\chi(r)-\frac{d}{\sqrt{r}}\triangle\chi(r).
\end{equation}
Due to the known behavior of $\phi$ the function $c(r,\theta,\varphi)$ is everywhere bounded and nonnegative. Last but not least we define the operator
\begin{equation}
\tilde{L}:=\triangle-\frac{c(\theta,\varphi,r)}{r^{2}}.
\end{equation}
With the help of these definitions, we arrive at
\begin{equation}
\tilde{L}h=f,
\end{equation}
which the function $h$ has to fulfill.
 Taking a closer look shows that $f\in C_{-5/2+\epsilon,5/2+\epsilon}^{0,\alpha}\left(\mathbb{R}^{3}\setminus\left\{0\right\}\right)$. But from Theorem (A.2) the map
\begin{equation}
\tilde{L}: C_{-1/2+\epsilon,1/2+\epsilon}^{2,\alpha}\left(\mathbb{R}^{3}\setminus\left\{0\right\}\right)\longrightarrow C_{-5/2+\epsilon,5/2+\epsilon}^{0,\alpha}\left(\mathbb{R}^{3}\setminus\left\{0\right\}\right)
 \end{equation}
is an isomorphism. It is obvious that the resulting initial data set has an asymptotically flat end as $r\rightarrow\infty$. To show that it also has an asymptotically cylindrical end, one uses the cylindrical metric
\begin{equation}
\tilde{h}_{ij}=\frac{d^{4}}{r^{2}}\delta_{ij},
\end{equation}
where $d$ denotes the solution of \eqref{equation5}.
\section{Conclusion}
This work was intended as an attempt to show how one can prove that initial data with similar properties like extreme Kerr black hole data exists. A proof of this was also given in \cite{dain} . But our method directly uses a suitable adapted version of the sub and supersolution theorem presented in \cite{choquet2}. There seems to be no need to start with wormhole-like configurations.
\section{Acknowledgments}
It is a pleasure to thank P.T.~Chru\'sciel for many useful discussions. This work was supported in part by Fonds zur F\"orderung der wissenschaftlichen Forschung in \"Osterreich, Projekt Nr. P20414-N16.\\

\appendix
\section{Mapping Properties}
\begin{lemma}
\label{lemma1}
Let $w$ satisfy the equation
\begin{equation}
\triangle w-\frac{c}{r^{2}}w=f,
\end{equation}
with $c\geq 0$ and $f\leq 0$. Suppose that $w\in C_{-1/2,1/2+\epsilon}^{2,\alpha}\left(\mathbb{R}^{3}\setminus \left\{0\right\}\right)$, where $0<\epsilon<\frac{1}{2}$. Then $w\geq 0$.
\end{lemma}
\emph{Proof.} We consider the following function
\begin{equation}
v:=r^{1/2+\epsilon/2}w.
\end{equation}
This function tends to $0$ at the origin and at infinity. Also $v\in C^{2,\alpha}\left(\mathbb{R}^{3}\setminus \left\{0\right\}\right)$.
We then compute
\begin{equation}\label{right}
\triangle v=r^{1/2+\epsilon/2}\triangle w+\frac{2\left(1/2+\epsilon/2\right)}{r}\nabla_{r}v+\left(\left(1/2+\epsilon/2\right)-\left(1/2+\epsilon/2\right)^{2}\right)\frac{1}{r^{2}}v.
\end{equation}
Suppose that at some point $v$ possesses a negative minimum. Then the r.h. side of (\ref{right}) is negative - a contradiction.
\begin{lemma}
Assume that $w\in C_{\delta_{1},\delta_{2}}^{0,\alpha}\left(\mathbb{R}^{3}\setminus\{0\}\right)$ and that $\tilde{w}\in C_{\tilde{\delta_{1}},\tilde{\delta_{2}}}^{0,\alpha}\left(\mathbb{R}^{3}\setminus\{0\}\right)$, then $w\tilde{w}\in C_{\delta_{1}+\tilde{\delta_{1}},\delta_{2}+\tilde{\delta_{2}}}^{0,\alpha}\left(\mathbb{R}^{3}\setminus\{0\}\right)$ and
\begin{equation}
\left\|w\tilde{w}\right\|_{C_{\delta_{1}+\tilde{\delta_{1}},\delta_{2}+\tilde{\delta_{2}}}^{0,\alpha}}\leq C\left\|w\right\|_{C_{\delta_{1},\delta_{2}}^{0,\alpha}}\left\|\tilde{w}\right\|_{C_{\tilde{\delta_{1}},\tilde{\delta_{2}}}^{0,\alpha}},
\end{equation}
for some constant $C>0$ independent of $w$ and $\tilde{w}$.
\end{lemma}
\emph{Proof.}
This is trivial except for the H\"older conditions. Concerning the latter
\begin{equation}
\frac{|w(x)\tilde{w}(x)-w(y)\tilde{w}(y)|}{|x-y|^{\alpha}}\leq\frac{|w(x)-w(y)|}{|x-y|^{\alpha}}|\tilde{w}(x)|+\frac{|\tilde{w}(x)-\tilde{w}(y)|}{|x-y|^{\alpha}}|w(y)|
\end{equation}
directly implies Lemma A.2.
\begin{lemma}
Assume that $\tilde{k}+\tilde{\alpha}<k+\alpha$ and that $\tilde{\delta_{1}}<\delta_{1},\tilde{\delta_{2}}<\delta_{2}$, then the embedding
\begin{equation}
I:C_{\delta_{1},\delta_{2}}^{k,\alpha}\left(\mathbb{R}^{3}\setminus\{0\}\right)\longrightarrow C_{\tilde{\delta_{1}},\tilde{\delta_{2}}}^{\tilde{k},\tilde{\alpha}}\left(\mathbb{R}^{3}\setminus\{0\}\right)
\end{equation}
is compact.
\end{lemma}
\emph{Proof.} This can be proven along the lines of \cite{chaljub}.
\begin{theo}
Let $c$ be a non-negative constant and $0<\epsilon<\frac{1}{2}$. Then,
\begin{eqnarray}
L: C_{-1/2,1/2+\epsilon}^{2,\alpha}\left(\mathbb{R}^{3}\setminus \left\{0\right\}\right)& \longrightarrow & C_{-5/2,5/2+\epsilon}^{0,\alpha}\left(\mathbb{R}^{3}\setminus \left\{0\right\}\right)\nonumber\\
 w & \longmapsto & \triangle w-\frac{c}{r^{2}}w
\end{eqnarray}
is an isomorphism.
\end{theo}
\emph{Proof.}
Injectivity follows from the previous lemma.
To prove surjectivity we use a domain decomposition method. We first look at the homogeneous problem. We denote the set of eigenfunctions of the Laplacian on $S^{2}$, i.e., the spherical harmonics, by $\phi_{j}, j\in\mathbb{N}$, that is
\begin{equation}
\triangle_{S^{2}}\phi_{j}=-\lambda_{j}\phi_{j}
\end{equation}
with
\begin{equation}
\lambda_{j} = j(j+1).
\end{equation}
If we project the operator on to the eigenspace spanned by $\phi_{j}$, we obtain the operator
\begin{equation}
L_{j}w:=\frac{\partial^{2}w}{\partial r^{2}}+\frac{2}{r}\frac{\partial w}{\partial r}-\frac{\lambda_{j}+c}{r^{2}}w.
\end{equation}
We may write the eigenfunction decomposition of $w$ as
\begin{equation}
w(r,\theta,\varphi)=\sum_{j\geq 0}w_{j}(r)\phi_{j}(\theta,\varphi).
\end{equation}
Any solution of $L_{j}w_{j}=0$ can be written as a linear combination of two linearly independent functions, which are given by
\begin{equation}
w_{j}^{+}:=r^{\gamma_{j}^{+}}
\end{equation}
and
\begin{equation}
w_{j}^{-}:=r^{\gamma_{j}^{-}},
\end{equation}
where we have set
\begin{equation}
\gamma_{j}^{\pm}=-\frac{1}{2}\pm\sqrt{\frac{1}{4}+\lambda_{j}+c}.
\end{equation}
The key observation is that the coefficients $\gamma_{j}^{\pm}$ determine all the possible asymptotic behaviors of the solutions of the homogeneous problem. As a next step we recall some properties of so-called \emph{Dirichlet-to-Neumann} maps. In our case these maps turn out to be fairly simple. First of all we introduce what we call the exterior Dirichlet-to-Neumann map. Let $\psi$ be any function in the space $C^{2,\alpha}\left(\partial B_{1}\right)$.
Given this boundary data, we define $v_{\psi}$ to be the unique solution of $Lv=0$ in $\mathbb{R}^{3}\setminus \bar{B}_{1}$ which belongs to $C_{\delta_{1},1/2+\epsilon}^{2,\alpha}$ and which satisfies $v=\psi$ on $\partial B_{1}$. It is well known that such a solution always exits. For the proof we refer the reader to \cite{chrusciel}. The exterior Dirichlet-to-Neumann map is now defined by
\begin{equation}
S\left(\psi\right):=\partial_{r}v_{\psi}|_{\partial B_{1}}.
\end{equation}
In order to have a better understanding of the operator $S$, let us consider the eigenfunction decomposition of $\psi\in C^{2,\alpha}\left(\partial B_{1}\right)$
\begin{equation}
\psi=\sum_{j\geq 0}\alpha_{j}\phi_{j}.
\end{equation}
Then we have the explicit formula
\begin{equation}
v_{\psi}=\sum_{j\geq 0}\alpha_{j}r^{\gamma_{j}^{-}}\phi_{j}.
\end{equation}
And thus, we obtain
\begin{equation}
\partial_{r}v_{\psi}|_{\partial B_{1}}=\sum_{j\geq 0}\gamma_{j}^{-}\alpha_{j}\phi_{j}.
\end{equation}
We now come to the definition of the interior Dirichlet-to-Neumann map. Again, let $\psi$ be any function in $C^{2,\alpha}\left(\partial B_{1}\right)$. This time, we define $w_{\psi}$ to be the unique solution of $Lw=0$ in $B_{1}$ which satisfies $w=\psi$ on $\partial B_{1}$ and belongs to $C_{-1/2,\delta_{2}}^{2,\alpha}$. Then the interior Dirichlet-to-Neumann map is defined by
\begin{equation}
T\left(\psi\right):=\partial_{r}w_{\psi}|_{\partial B_{1}}.
\end{equation}
In our simple case, we even have an explicit representation of $T$. Let $\psi$ be decomposed in terms of the eigenfunctions again. Then, $w_{\psi}$ is explicitly given by
\begin{equation}
w_{\psi}=\sum_{j\geq 0}\alpha_{j}r^{\gamma_{j}^{+}}\phi_{j}.
\end{equation}
Hence
\begin{equation}
\partial_{r}w_{\psi}|_{\partial B_{1}}=\sum_{j\geq 0}\gamma_{j}^{+}\alpha_{j}\phi_{j}.
\end{equation}
We claim that
\begin{equation}
S-T:C^{2,\alpha}\left(\partial B_{1}\right)\longrightarrow C^{1,\alpha}\left(\partial B_{1}\right),
\end{equation}
is an isomorphism. Indeed, in terms of the eigenfunctions $\phi_{j}$ the map is given by
\begin{equation}
\sum_{j\geq 0}\alpha_{j}\phi_{j}\longrightarrow \sum_{j\geq 0}\left(\gamma_{j}^{-}-\gamma_{j}^{+}\right)\alpha_{j}\phi_{j}.
\end{equation}
Now notice that $\gamma_{j}^{-}-\gamma_{j}^{+}\neq 0$. From this our claim easily follows. We proceed to construct a right inverse of $L$. For all $f\in C_{-5/2,5/2+\epsilon}^{0,\alpha}$, we define $w_{ext}$ to be the unique solution of
\begin{equation}
Lw_{ext}=f
\end{equation}
in $\mathbb{R}^{3}\setminus\bar{B}_{1}$ that vanishes on $\partial B_{1}$ and belongs to $C_{\delta_{1},1/2+\epsilon}^{2,\alpha}$. The existence of $w_{ext}$ is again guaranteed \cite{chrusciel}. We also define $w_{int}$ to be the solution of
\begin{equation}
Lw_{int}=f
\end{equation}
in $B_{1}\setminus\left\{0\right\}$ that vanishes on $\partial B_{1}$ and belongs to $C_{-1/2,\delta_{2}}^{2,\alpha}$.
Now, for existence of $w_{int}$, we consider the eigenfunction decomposition of $f$
\begin{equation}
f=\sum_{j\geq0}f_{j}\phi_{j},
\end{equation}
and look for $w_{int}$ of the form
\begin{equation}
w_{int}=\sum_{j\geq0}w_{j}\phi_{j}.
\end{equation}
Hence we have to solve the ordinary differential equations $L_{j}w_{j}=f_{j}$ in $(0,1]$, with the boundary condition $w_{j}\left(1\right)=0$. It is easy to show that the following formula for $w_{j}$ satisfies the equations. Namely,
\begin{equation}
w_{j}=-r^{\gamma_{j}^{+}}\int_{r}^{1}s^{-2-2\gamma_{j}^{+}}\int_{0}^{s}t^{2+\gamma_{j}^{+}}f_{j}(t)dtds.
\end{equation}
Notice that this expression is well defined since $0<\gamma_{j}^{+}$. Furthermore, we can estimate
\begin{equation}
\sup_{\left(0,1\right]}\left|r^{1/2}w_{j}\right|\leq c_{j} \sup_{\left(0,1\right]}\left|r^{5/2}f_{j}\right|,
\end{equation}
for some constant $c_{j}>0$. Hence we can assume that there exists some constant $c_{J}>0$, which is independent of $f$ but may a priori depend on $J$, such that
\begin{equation}
\sup_{B_{1}\setminus\left\{0\right\}}\left|r^{1/2}\sum_{j=0}^{J}w_{j}\phi_{j}\right|\leq c_{J}\sup_{B_{1}\setminus\left\{0\right\}}\left|r^{5/2}f\right|.
\end{equation}
It remains to prove that the constant $c_{J}$ does not depend on $J$. Once this is known, we can pass to the limit $J\to\infty$ and easily obtain a solution of our problem.  Along the lines of \cite{pacard} one can argue by contradiction and assume that the claim is not true. This assumption easily conflicts with the injectivity of $L$ on certain subspaces of functions. The fact that $w_{int}\in C_{-1/2,\delta_{2}}^{2,\alpha}$ then follows by using rescaled Schauder estimates. Note that our estimates imply
\begin{equation}
\sup_{B_{1}\setminus\left\{0\right\}}\left|r^{1/2}w_{int}\right|\leq c \sup_{B_{1}\setminus\left\{0\right\}}\left|r^{5/2}f\right|.
\end{equation}
The reader is referred to \cite{pacard} for further details. The final step of the construction is looking for a solution of
\begin{equation}
Lw_{ker}=0
\end{equation}
in $\mathbb{R}^{3}\setminus\partial B_{1}$, which is continuous in $\mathbb{R}^{3}\setminus\left\{0\right\}$ and possesses the right asymptotic limits. In addition, we want to be able to choose $w_{ker}$ in such a way that the function
\begin{equation}
w:=\left\{\begin{array}{ll} w_{ext}+w_{ker} & \textrm{in $\mathbb{R}^{3}\setminus \bar{B}_{1}$}\\ & \\ w_{int}+w_{ker} & \textrm{in $B_{1}\setminus\{0\}$}\end{array}\right.
\end{equation}
is also differentiable in all $\mathbb{R}^{3}\setminus\left\{0\right\}$. Assuming we have already constructed $w_{ker}$, we see that
\begin{equation}
Lw=f,
\end{equation}
in $\mathbb{R}^{3}\setminus\left\{0\right\}$ and that we have found a right inverse for $L$ with the desired properties. In order to build $w_{ker}$, we must check that we can find a solution $Lw_{ker}=0$, which is continuous across $\partial B_{1}$ and for which the discontinuity of $\partial_{r}w_{ker}$ through $\partial B_{1}$ is equal to $\left(\partial_{r}w_{ext}-\partial_{r}w_{int}\right)|_{\partial B_{1}}$. Remember that solutions of $Lw_{ker}=0$ are parameterized by their values on $\partial B_{1}$. So we are done when we are able to show that there exists a function $\psi\in C^{2,\alpha}\left(\partial B_{1}\right)$ which is a solution of the equation
\begin{equation}
\left(S-T\right)\left(\psi\right)=-\left(\partial_{r}w_{ext}-\partial_{r}w_{int}\right)|_{\partial B_{1}}.
\end{equation}
But we already know that the
\begin{equation}
S-T:C^{2,\alpha}\left(\partial B_{1}\right)\longrightarrow C^{1,\alpha}\left(\partial B_{1}\right)
\end{equation}
is an isomorphism. The desired properties of $w_{ker}$ then follow from the constructions of the Dirichlet-to-Neumann maps. The proof of surjectivity of $L$ is therefore complete.
\begin{theo}
Let $\tilde{c}$ be a nonnegative smooth bounded function in $\mathbb{R}^{3}\setminus\left\{0\right\}$ and $0<\epsilon<\frac{1}{2}$. Then,
\begin{eqnarray}
\tilde{L}: C_{-1/2+\epsilon,1/2+\epsilon}^{2,\alpha}\left(\mathbb{R}^{3}\setminus \left\{0\right\}\right)& \longrightarrow & C_{-5/2+\epsilon,5/2+\epsilon}^{0,\alpha}\left(\mathbb{R}^{3}\setminus \left\{0\right\}\right)\nonumber\\
 w & \longmapsto & \triangle w-\frac{\tilde{c}}{r^{2}}w
\end{eqnarray}
is an isomorphism.
\end{theo}
\emph{Proof.} First of all we want to point out that the different weights (compare this theorem with the previous theorem) at the origin are only chosen because of the applications that we have in mind. In fact the operators $L$ and $\tilde{L}$ are isomorphisms between the same weighted function spaces. For our proofs to work, it is only important that the weights which belong to the function spaces in the domain of the operators fulfill certain assumptions, namely $\delta_{1},-\delta_{2}\in (\gamma_{0}^{-},\gamma_{0}^{+})$ and that $-\delta_{1}<\delta_{2}$. Most likely proofs exist without the need of the second assumption.

Again a similar argument like in the previous theorem can be used to show that the operator is injective. To prove surjectivity we can use the fact that we have already shown that when $\tilde{c}$ is a nonnegative constant the mapping is surjective. This is enough to define a global subsolution and also a global supersolution which in turn implies surjectivity of the more general mapping. Let $f$ be any function that belongs to $C_{-5/2+\epsilon,5/2+\epsilon}^{0,\alpha}\left(\mathbb{R}^{3}\setminus \left\{0\right\}\right)$. We show that there exist $w_{-},w_{+}\in C_{-1/2+\epsilon,1/2+\epsilon}^{2,\alpha}\left(\mathbb{R}^{3}\setminus \left\{0\right\}\right)$ such that
\begin{equation}
\tilde{L}w_{-}\geq f,
\end{equation}
\begin{equation}
\tilde{L}w_{+}\leq f
\end{equation}
and
\begin{equation}
w_{-}\leq w_{+}
\end{equation}
pointwise. Let $c$ be a constant with the following property:
\begin{equation}
0\leq c\leq\sup_{\mathbb{R}^{3}} c(r,\theta,\varphi).
\end{equation}
Now define $w_{-}$ to be the unique solution to
\begin{equation}
\label{equation9}
\left(\triangle-\frac{c}{r^{2}}\right)w_{-}=f_{1},
\end{equation}
which belongs to $C_{-1/2+\epsilon,1/2+\epsilon}^{2,\alpha}\left(\mathbb{R}^{3}\setminus \left\{0\right\}\right)$ and where $f_{1}$ denotes a function with the following properties:
\begin{itemize}
\item[(1)] $f_{1}\in C_{-5/2+\epsilon,5/2+\epsilon}^{0,\alpha}\left(\mathbb{R}^{3}\setminus \left\{0\right\}\right),$
\item[(2)] $f_{1}\geq f,$
\item[(2)] $f_{1}\geq 0.$
\end{itemize}
Similarly we define $w_{+}$ to be the unique solution (see Theorem A.1) to
\begin{equation}
\label{equation10}
\left(\triangle-\frac{c}{r^{2}}\right)w_{-}=f_{2},
\end{equation}
which belongs to $C_{-1/2+\epsilon,1/2+\epsilon}^{2,\alpha}\left(\mathbb{R}^{3}\setminus \left\{0\right\}\right)$ and where $f_{2}$ denotes a function with the following properties:
\begin{itemize}
\item[(1)] $f_{2}\in C_{-5/2+\epsilon,5/2+\epsilon}^{0,\alpha}\left(\mathbb{R}^{3}\setminus \left\{0\right\}\right),$
\item[(2)] $f_{2}\leq f,$
\item[(2)] $f_{1}\leq 0.$
\end{itemize}
Because of the previous theorem we know that solutions to \eqref{equation9} and \eqref{equation10} always exist. One can easily show that these solutions possess the desired properties, e.g. $w_{-}\leq w_{+}$ follows because our assumptions imply that $w_{-}\leq 0$ and $w_{+}\geq 0$. Now the existence of global functions $w_{-}$ and $w_{+}$ that belong to the right space is always enough to construct a right inverse for $\tilde{L}$ by standard methods. For example the solvability of the Dirichlet problem in arbitrary compact domains can be proven by using the method of continuity. $w_{-}$ and $w_{+}$ imply bounds which then can be used to extend the solution on $\mathbb{R}^{3}\setminus \left\{0\right\}$.

\end{document}